%% file: FrequencyDependentPowerModelingofCMOSTransmittersforWNoCArchitectures.tex
\newcommand{\CLASSINPUTtoptextmargin}{19mm}%
\newcommand{\CLASSINPUTbottomtextmargin}{43mm}%
\newcommand{\CLASSINPUTinnersidemargin}{12.9mm}%
\newcommand{\CLASSINPUToutersidemargin}{12.9mm}%
\documentclass[conference,10pt,a4paper]{IEEEtran}
\usepackage{amsmath}
\usepackage{times}
\usepackage{graphicx}
\usepackage{multirow}
\usepackage[none]{hyphenat}
\usepackage{float}
\usepackage{subfig}
\usepackage{iftex}
\ifPDFTeX%
\usepackage{t1enc}
\usepackage{times}
\fi%
\ifXeTeX%
\usepackage{fontspec}
\fi%
\ifLuaTeX%
\usepackage{fontspec}
\fi%
\input{modify_IEEEtran}

\begin{document}
\raggedbottom
%

\title{Frequency-Dependent Power Consumption Modeling of CMOS Transmitters for WNoC Architectures}
%
%
\author{%
\IEEEauthorblockN{%
Mohammad Shahmoradi\EuMWauthorrefmark{*}, 
Korkut Kaan Tokgöz\EuMWauthorrefmark{**}, 
Eduard Alarcón\EuMWauthorrefmark{*}, 
Sergi Abadal\EuMWauthorrefmark{*}
}%
\IEEEauthorblockA{\EuMWauthorrefmark{*}Universitat Polit\`ecnica de Catalunya, Barcelona, Spain}%
\IEEEauthorblockA{\EuMWauthorrefmark{**}Sabanci University, Istanbul, Turkey}%
}

\maketitle

%

\begin{abstract}
Wireless Network-on-Chip (WNoC) systems, which wirelessly interconnect the chips of a computing system, have been proposed as a complement to existing chip-to-chip wired links. However, their feasibility depends on the availability of custom-designed high-speed, tiny, ultra-efficient transceivers. This represents a challenge due to the tradeoffs between bandwidth, area, and energy efficiency that are found as frequency increases, which suggests that there is an optimal frequency region. To aid in the search for such an optimal design point, this paper presents a behavioral model that quantifies the expected power consumption of oscillators, mixers, and power amplifiers as a function of frequency. The model is built on extensive surveys of the respective sub-blocks, all based on experimental data. By putting together the models of the three sub-blocks, a comprehensive power model is obtained, which will aid in selecting the optimal operating frequency for WNoC systems.

\end{abstract}


\section{Introduction}
Advancements in integrated circuit design, particularly in multi-core processors, have enabled the development of Wireless Network-on-Chip (WNoC) communication, which has garnered considerable research interest in recent years~\cite{ghasemi2025ultra,paudel2024design,yazdanpanah2023low} due to its potential to overcome the limitations of wired on-chip communication. WNoC leverages wireless high-speed links to enable low-latency data transfer between distant cores, offering a reasonable alternative to wired solutions. Additionally, seamless integration with high-frequency CMOS technologies, essential for implementing wireless transceivers on-chip, enhances appeal.

Despite its advantages, WNoC encounters substantial obstacles related to dimensional constraints and power use, which limit the design and prototyping processes. Implementing and measuring prototypes remains a crucial method for assessing design performance; however, this is an expensive and time-consuming endeavor, especially for WNoC systems that demand precise design considerations. To mitigate these issues, robust simulation and modeling methodologies that minimize the dependency on costly prototyping are required. In this light, behavioral modeling utilizing the specifications of fabricated prototypes offers a framework for system evaluation that is both time-efficient and cost-effective.

\begin{figure}[t]
\centering
\includegraphics[width=85mm]{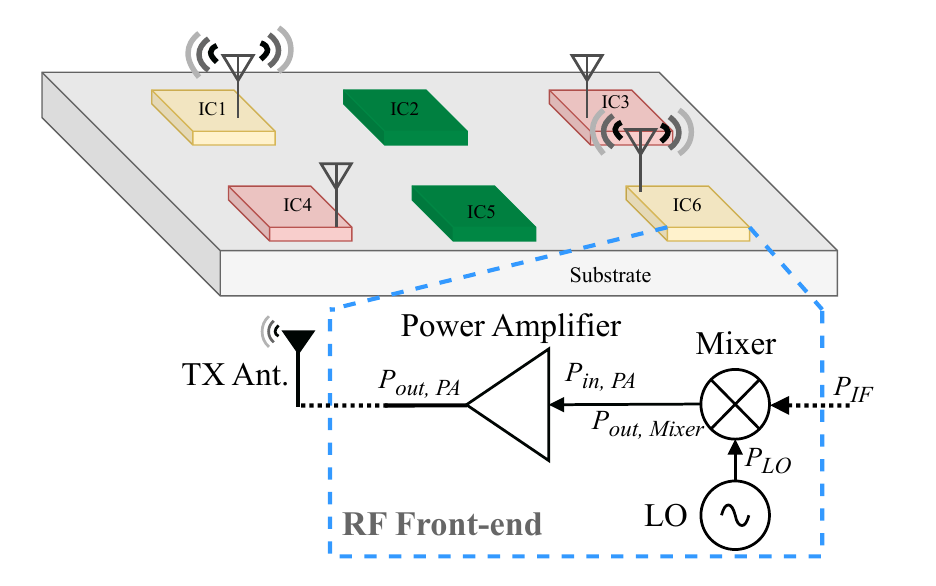}  
\caption{Wireless Network-on-Chip (WNoC) architecture in a multi-chip system. The inset shows a simplified RF front-end for the TX.}
\label{fig: WNoC architecture_sub_block}
\vspace{-\baselineskip}
\end{figure}

Recent advancements in behavioral modeling have primarily focused on mm-wave and terahertz (THz) systems, enabling high data rates in next-generation wireless technologies~\cite{hietanen2025technology}. While these models are well-suited for long-range wireless communication, WNoC systems operate over ultra-short distances and offer greater flexibility in frequency band selection. This distinction presents an opportunity to develop frequency-dependent behavioral models tailored to WNoC architectures.

Existing behavioral models often overlook system-level interactions between sub-blocks~\cite{abadal2022graphene}, which are essential for analyzing power consumption trends in WNoC architectures. To bridge this gap, a comprehensive modeling framework is required, one that not only covers RF front-end components, the most power-hungry part of transceivers, but also enhances feasibility by applying fabricated prototypes like those documented in~\cite{wang2020power}. Additionally, for WNoC designs, making certain assumptions about varying power levels could support alignment with their potential conditions.

Therefore, this research seeks to provide a breakdown of power consumption across RF front-end sub-blocks, including the PA, mixer, and oscillator, with respect to operating frequency for WNoC applications (Fig.~\ref{fig: WNoC architecture_sub_block}). The proposed model also adaptively calculates power consumption based on the required \( P_{\text{IF}} \) and \( P_{\text{out,TX}} \) for modulated data in the CMOS TX chain. Ultimately, the model quantifies the contribution of each TX chain component and recommends the optimal operating frequency to improve power efficiency.

The remainder of this paper is organized as follows: Section~\ref{Section: Methodology} describes the methodology for developing power models of the front-end components. Section~\ref{Section: Results} presents an analysis of the model’s power consumption performance. Finally, Section~\ref{Section: Conclusion} summarizes the key findings of this work.


\section{Methodology}\label{Section: Methodology}


WNoC architectures are constrained by power consumption while benefiting from the flexibility of operating in frequencies that do not interfere with external wireless communication. Since WNoC operates within an enclosed physical volume (i.e., inside a chip), it does not face EMI/EMC restrictions imposed on traditional wireless systems. Therefore, this paper develops frequency-dependent power models for the RF front-end components of a CMOS TX in WNoC architecture, focusing on the power amplifier (PA), oscillator, and mixer, operating at frequencies up to the THz range. These models provide a framework for characterizing dissipated power trends and enabling time-efficient design strategies.

To ensure feasibility, prototype surveys of front-end components were utilized, encompassing critical PA and oscillator characteristics from~\cite{wang2020power} and mixer specifications from~\cite{zhang2022optimal}. Individual models are systematically integrated to quantify the dissipated power of each component across operational scenarios in WNoC architectures.

The methodology employs survey data to formulate the model framework, providing a reference for defining structural assumptions. However, slight deviations are inevitable since the surveyed prototypes were not designed for WNoC applications. Key parameters are initialized based on the literature to align the model with WNoC requirements.

\subsection{Modeling of Power consumption in Power Amplifiers}\label{subsection: PA_dc_Power}

As the PA is the most power-hungry component in RF front-ends, it was prioritized in the modeling process. Achieving a precise model for power consumption based on the Power Added Efficiency (PAE) is challenging due to dependencies on factors such as biasing class, technology, and operating conditions~\cite{egan2004practical}. However, the frequency-scalable model developed in this study for the PA (\(\mathrm{P_{DC,~PA}}\)) provides sufficient accuracy under the initial assumptions tailored for WNoC architectures. \(\mathrm{P_{DC,~PA}}\) is determined using Eq.~\eqref{eqn: Pdc},
\begin{equation}
P_{\mathrm{DC,~PA}}^{\mathrm{mW}}(f) = \frac{P_{\mathrm{out}}^{\mathrm{mW}} - P_{\mathrm{in}}^{\mathrm{mW}}}{0.01 \cdot \mathrm{PAE}(f)},
\label{eqn: Pdc}
\end{equation}
where $f$ denotes the operating frequency. Given that the majority of prototype modulation schemes in~\cite{wang2020power} employ Quadrature Amplitude Modulation (QAM), ensuring PA operation within its linear region is essential. Additionally, in short-range on-chip communication systems, high-power PAs are generally unnecessary, making low-power versions more suitable for WNoC systems. To maintain linearity while operating within the required power range for WNoC systems, the output power (\( \mathrm{P_{out}} \))  is assumed to be 0 dBm for the modeled PAs.
On the other hand, to analyze PA gain characteristics, input power ($\mathrm{P_{in}}$) levels from $-15$ to $0$ dBm were chosen, covering typical PAs in~\cite{wang2020power} and scenarios where a PA is unnecessary (0 dB gain).

\subsection{Modeling of Power consumption in Oscillators}\label{subsection: OSc_Behavioral_Modeling}

In this study, fundamental oscillators are selected for their superior power efficiency, a key factor in addressing energy constraints in high-frequency applications. Unlike harmonic oscillators, which rely on complex, power-intensive circuitry, fundamental ones offer an efficient and scalable solution without performance trade-offs. This makes them ideal for applications where minimizing power consumption is critical.

The oscillator power modeling approach in this work follows the PA methodology outlined in Section~\ref{subsection: PA_dc_Power}, with initial insights drawn from the survey on THz oscillators~\cite{wang2020power}. For additional surveys, we focused specifically on fundamental oscillators operating up to 272 GHz~\cite{landsberg2013240}, ensuring a comprehensive representation of oscillator behavior across the target frequency spectrum. Finally, the power consumption of oscillators is characterized by the DC-to-RF efficiency ratio, which considers the dissipated power defined in Eq.~\eqref{eqn: Osc_dcRF}.

\begin{equation}
\text{DC-to-RF Efficiency(\textit{f})} = \frac{P_{\text{RF, Osc.}}}{P_{\text{DC, Osc.}(f)}}.
\label{eqn: Osc_dcRF}
\end{equation}

Based on the assumptions in Section~\ref{subsection: PA_dc_Power}, the same range was applied to $\mathrm{P_{RF, Osc.}}$. These values were selected to represent diverse operating conditions and enable a robust analysis of oscillator performance.

\subsection{Modeling of Power consumption in Mixers}
Mixers are primarily classified as passive and active, with most implementations based on Gilbert cells~\cite{razavi2021fundamentals}. According to WNoC requirements, passive mixers are preferred for our modeling due to their lower power requirements and enhanced energy efficiency. This choice directly supports the goal of optimizing power efficiency in WNoC applications. Eventually, for the power modeling of mixers, we used the conversion gain (CG) or conversion loss (CL), 

\begin{equation}
\text{CG (dB)} = 10 \cdot \log_{10} \left( \frac{P_{\text{RF, out}}}{P_{\text{IF, in}}} \right) = - \text{CL (dB)}.
\label{eqn: Mix_CG}
\end{equation}

To develop a power model for passive mixers, a $\mathrm{P_{IF, in}}$ of 3 dBm was assumed, applying the \( \mathrm{P_{in}} \) range from Section~\ref{subsection: PA_dc_Power} to \(\mathrm{P_{RF,~out}}\). The mixer survey~\cite{zhang2022optimal} benchmarked the ratio of reported linear conversion gain to \(\mathrm{P_{dc}}\). Using best-in-class data and these assumptions, the mixer's power consumption was modeled as a function of operating frequency.

\section{Results}\label{Section: Results}

This section presents modeling results based on published data \cite{zhang2022optimal,wang2020power}, integrating frequency-dependent power models of RF front-end sub-blocks into a comprehensive TX chain model optimized for WNoC architectures.

\subsection{Behavioral Power Model of the Power Amplifier}\label{subsection: PA_Result}

\begin{figure}[t]
\centering
\includegraphics[width=85mm]{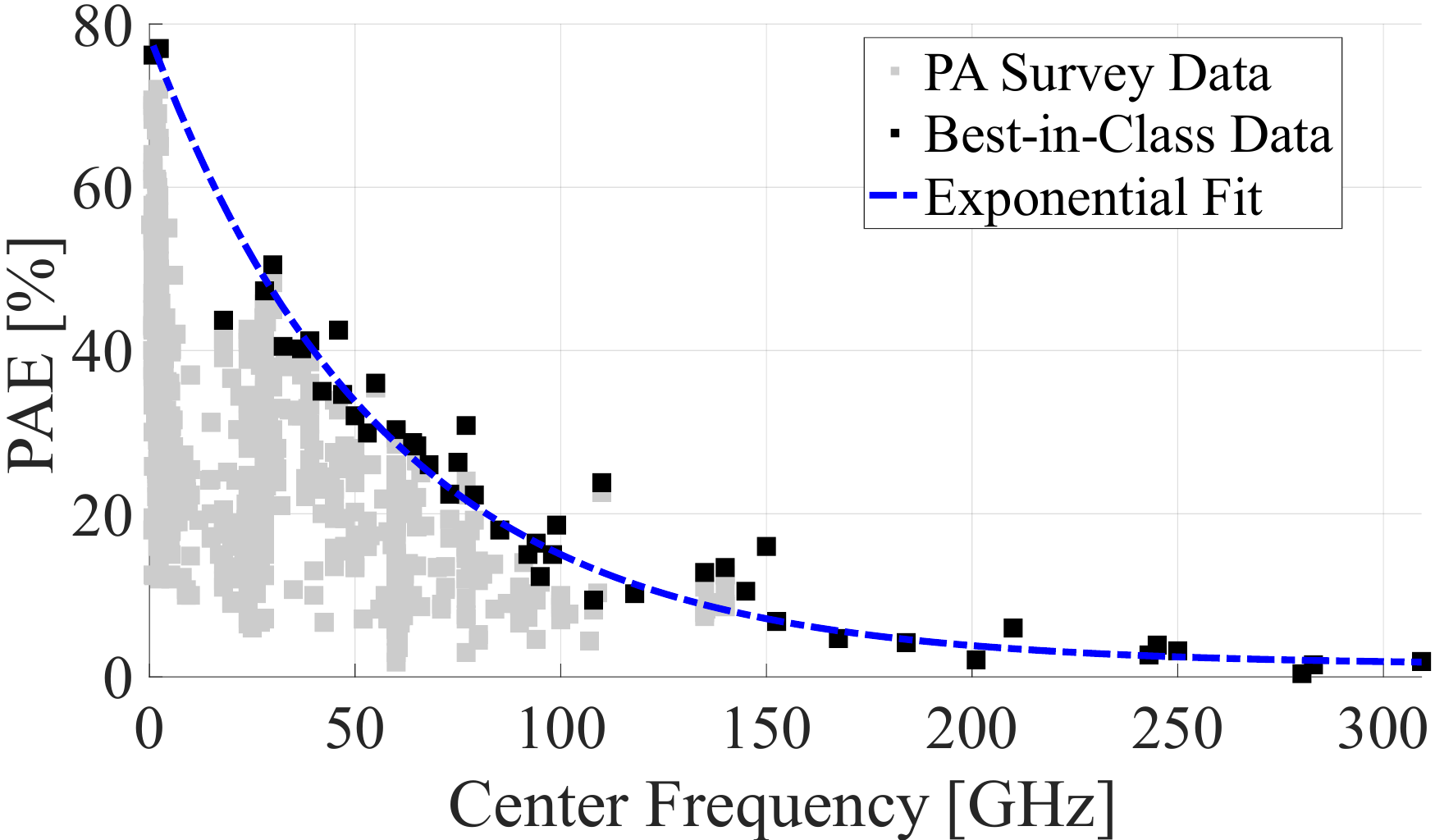} 
\includegraphics[width=85mm]{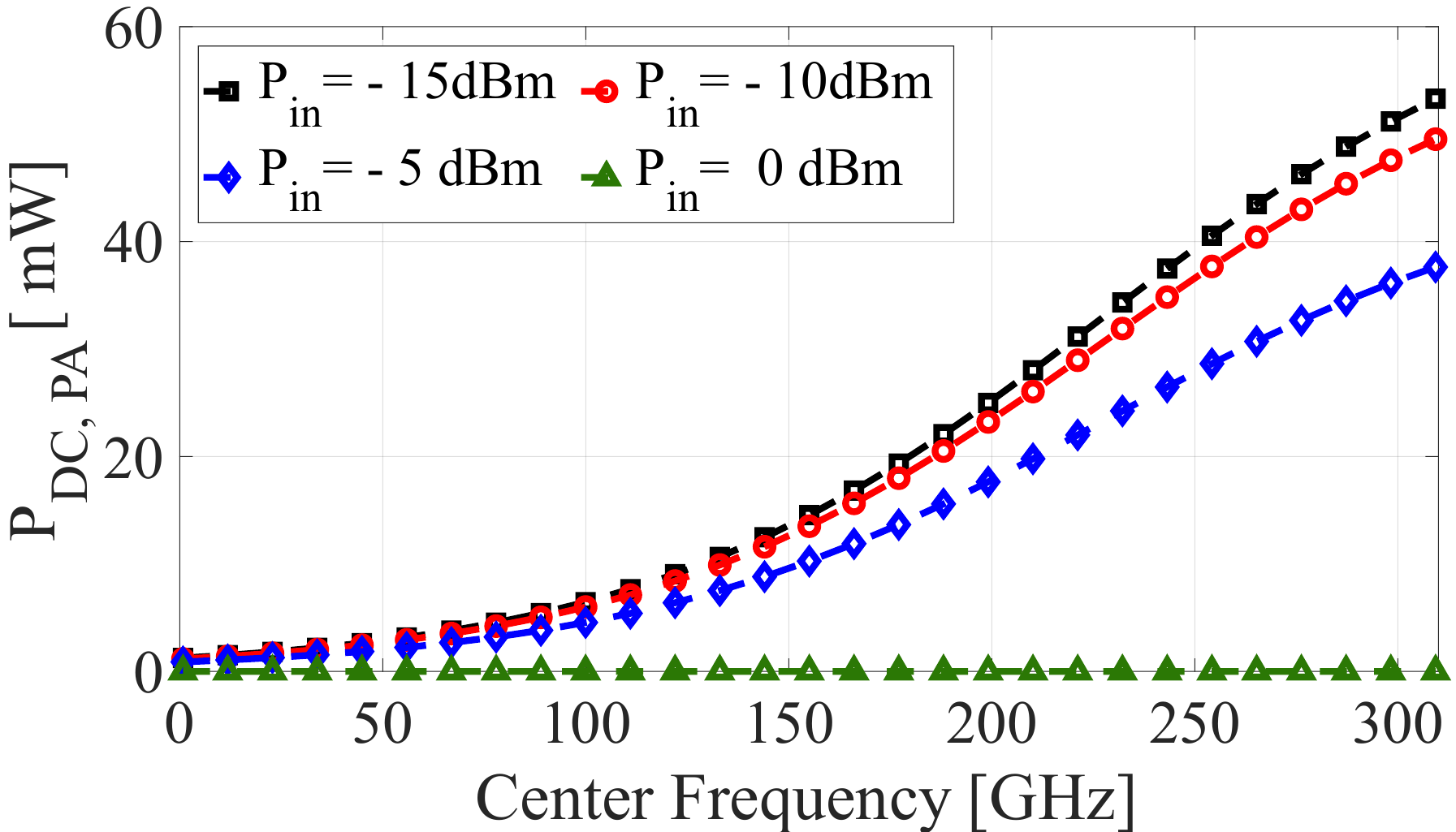} 
\caption{Frequency-dependent PAE model based on best-in-class data from~\cite{wang2020power}(top) and resulting \(\mathrm{P_{DC, PA}}\) for multiple \( \mathrm{P_{in}} \) configurations (bottom).}
\label{fig: PAE_dc_Power_PA_Survey}
\vspace{-\baselineskip}
\end{figure}

We plotted the PAE as a function of operating frequency and identified the best-in-class data points (black squares in Fig.~\ref{fig: PAE_dc_Power_PA_Survey}). By applying an exponential fit (Fig.~\ref{fig: PAE_dc_Power_PA_Survey}, top), we formulated a frequency-dependent equation to characterize the efficiency trend of PAs, achieving an R-squared equal to 0.6, over the frequency range $0.9 \leq f \leq 309.3~\text{GHz}$.

To extract the dissipated power trend for the PA sub-block, we followed the assumptions in Section~\ref{subsection: PA_dc_Power}. Assuming $\mathrm{P_{out,~PA}} = 0$ dBm for low-power PAs and considering four values of $\mathrm{P_{in,~PA}}$ = [-15, -10, -5, 0] dBm, we applied Eq.~\eqref{eqn: Pdc} to determine $\mathrm{P_{DC,~PA}}$ (Fig.~\ref{fig: PAE_dc_Power_PA_Survey}, bottom).

As shown in Fig.~\ref{fig: PAE_dc_Power_PA_Survey} (bottom), \(\mathrm{P_{DC}}\) exhibits an exponential increase with rising operating frequency. At high frequencies, achieving a gain of 15 dB can result in PA power consumption reaching several tens of milliwatts. In contrast, for a scenario with 0 dB gain, the PA is unnecessary in WNoC applications. Moreover, in this model, the \(\mathrm{P_{in}}\) and \(\mathrm{P_{out}}\) of PAs can be customized to satisfy particular requirements.

\subsection{Behavioral Power Model of the Fundamental Oscillator}\label{sebseciton: OSc_Results}

\begin{figure}[t]
\centering
\includegraphics[width=85mm]{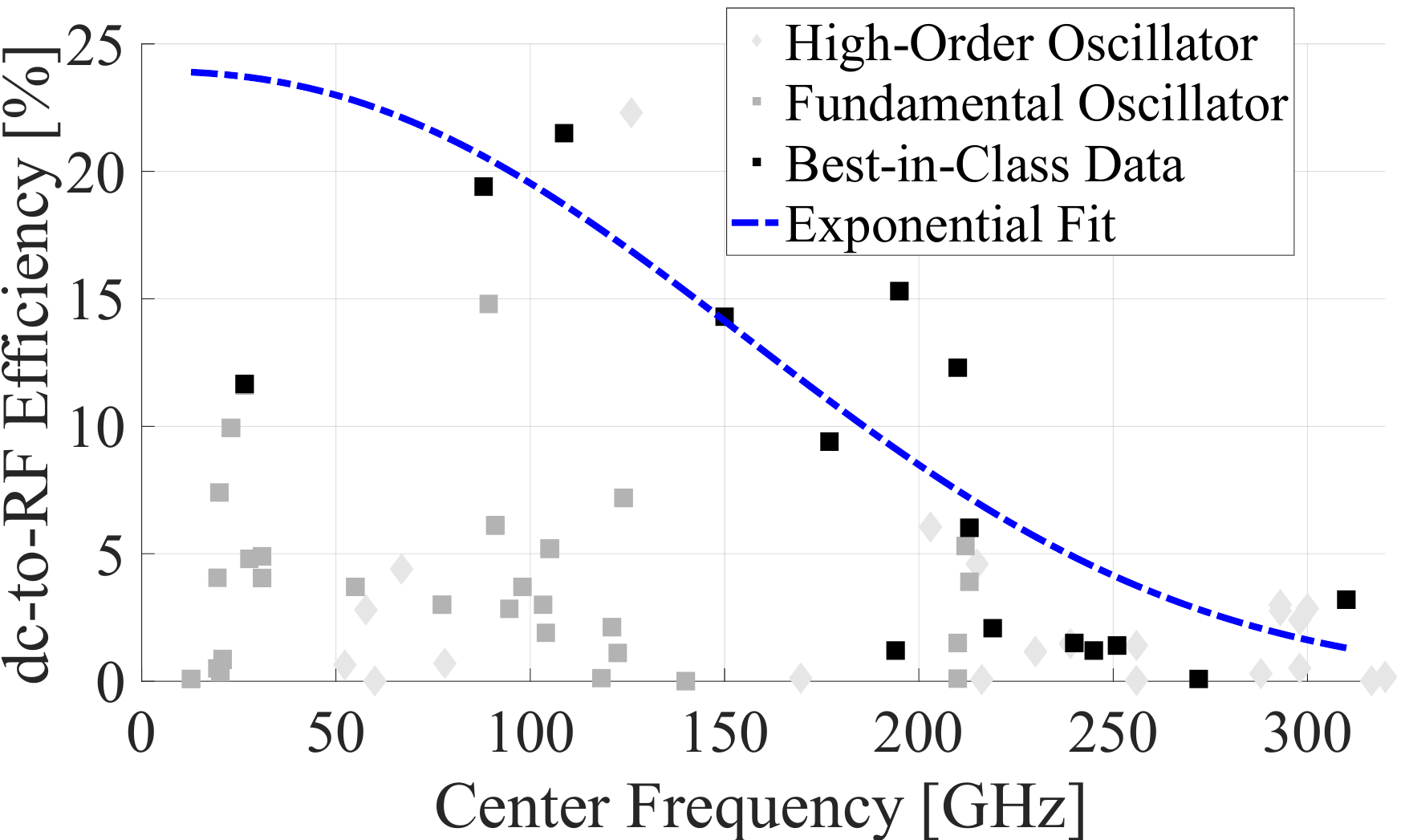} 
\includegraphics[width=85mm]{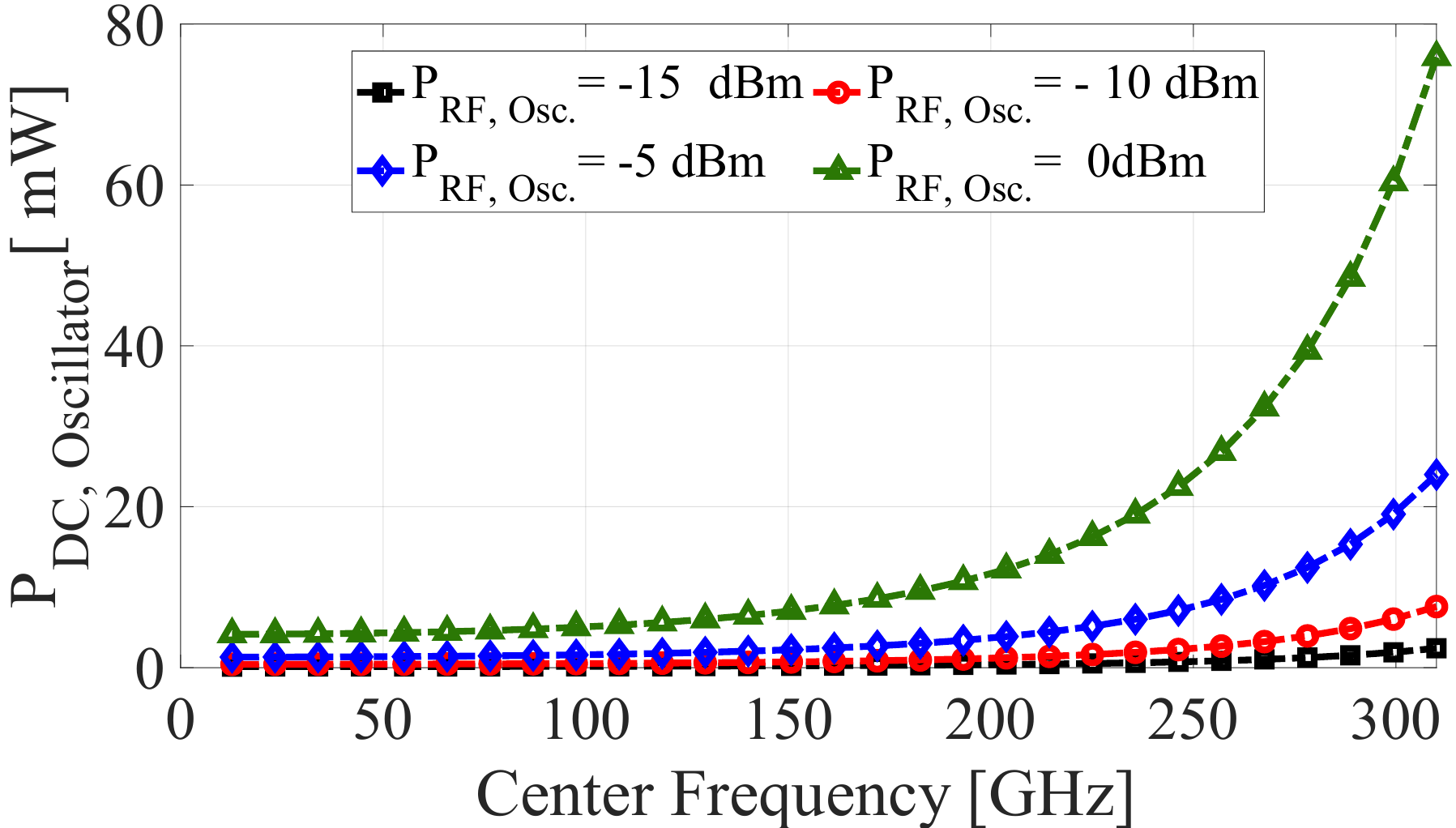} 
\caption{Frequency-dependent oscillator model: DC-to-RF efficiency (top) and $\mathrm{P_{DC, Osc}}$ for varying $\mathrm{P_{RF,~out}}$ (bottom).}
\label{fig: dc_Power_Oscillator_Survey}
\vspace{-\baselineskip}
\end{figure}

Following the methodology in Section~\ref{subsection: PA_Result}, regression analysis was performed to derive an exponential trend for fundamental oscillators based on best-in-class data (black squares in Fig.~\ref{fig: dc_Power_Oscillator_Survey}). This approach ensured the extraction of an optimistic model for $12.7 \leq f \leq 310~\text{GHz}$ (Fig.~\ref{fig: dc_Power_Oscillator_Survey}, top) and resulted in an R-squared value close to 0.7.

Utilizing the DC-to-RF efficiency (Eq.~\eqref{eqn: Osc_dcRF}), we observed the pattern of $\mathrm{P_{DC,~Osc.}}$, as shown in Fig.~\ref{fig: dc_Power_Oscillator_Survey}. Our findings suggest that fundamental oscillators under 200~GHz within WNoC systems use less than 10~mW of DC power, ideal for low-power uses. In contrast, their power consumption rises markedly in THz ranges, reducing efficiency and increasing costs. Thus, oscillators can be energy-demanding, and in low-power PA scenarios, their energy use might match or surpass that of a PA, indicating a need for design optimization in high-frequency settings.

\subsection{Behavioral Power Model of the Mixer}\label{subsection: Mixer_dc_Power}

\begin{figure}[t]
\centering
\includegraphics[width=85mm]{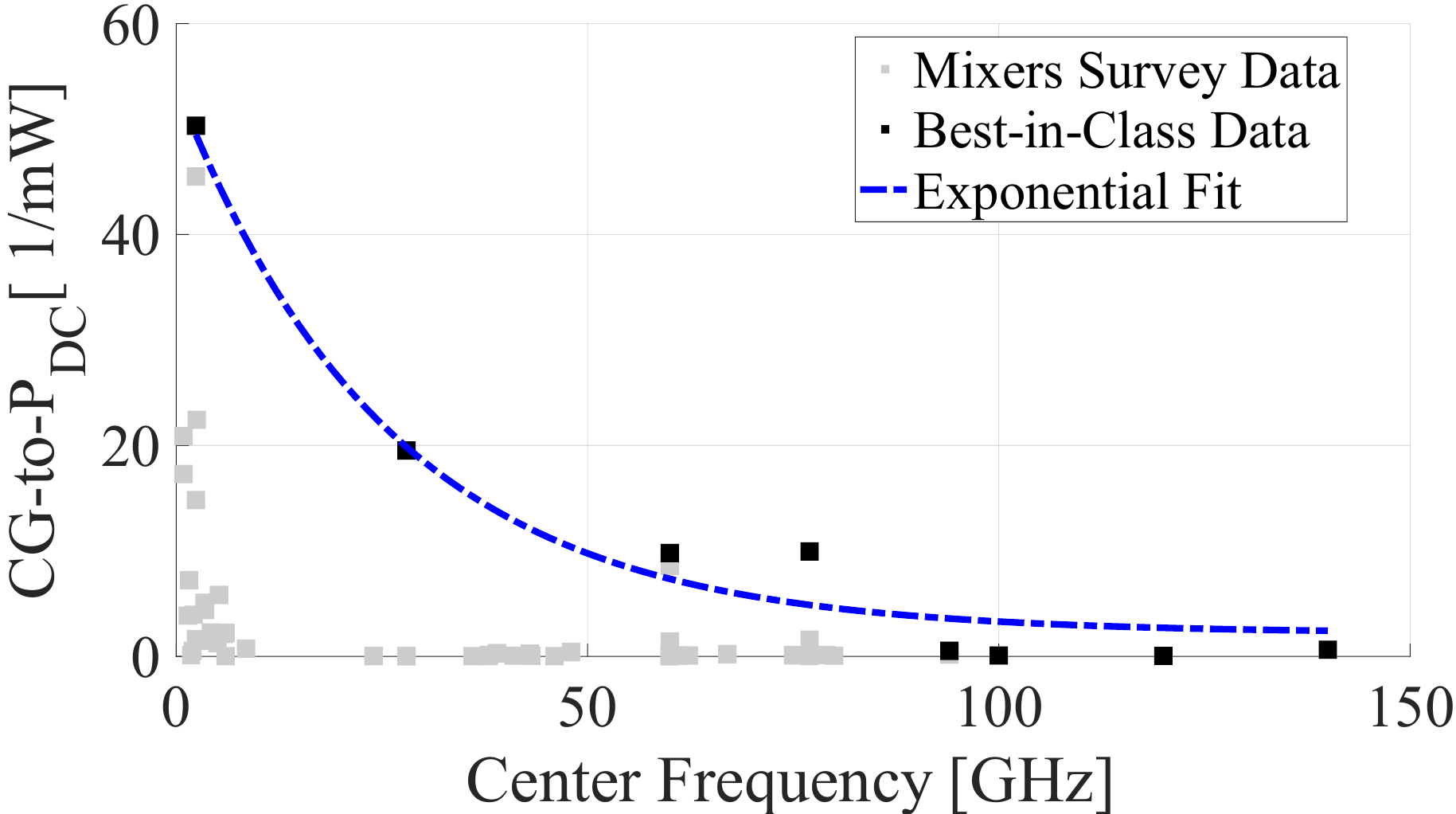} 
\includegraphics[width=85mm]{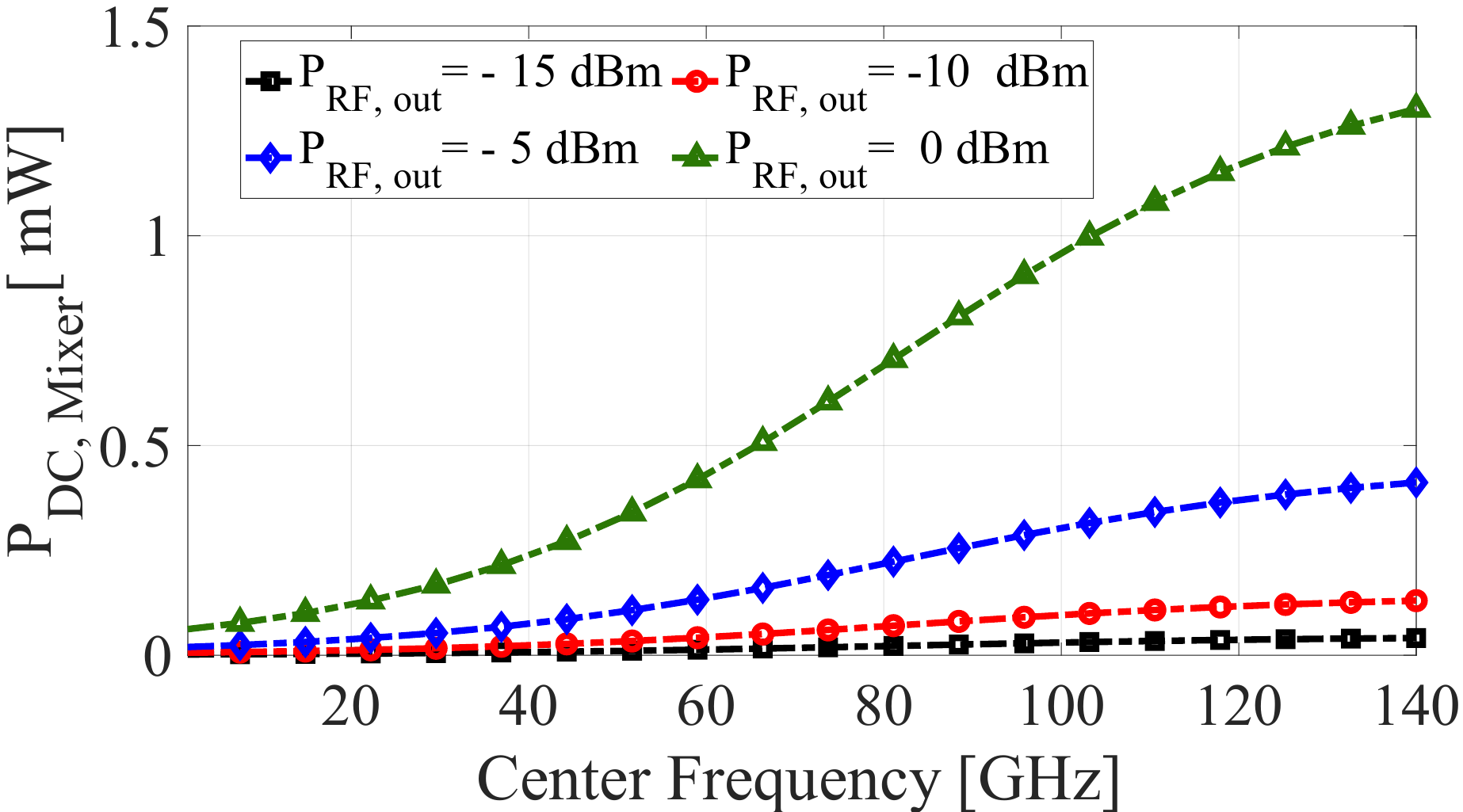} 
\caption{Mixer power consumption model as a function of frequency (top) and derived model for \(\mathrm{P_{DC,~Mixer}}\) (bottom).}
\label{fig: Mixer_CG_Over_Pdc}
\vspace{-\baselineskip}
\end{figure}

The determination of the CG-to-\(\mathrm{P_{DC}}\) ratio is derived from the analysis in~\cite{razavi2021fundamentals}, which shows that CG (or CL) is proportional to the product of \(R_L\) and \(\mathrm{P_{DC}}\), divided by \(V_{DD}\). The CG-to-\(\mathrm{P_{DC}}\) was examined over a broad frequency range, aligned with best-in-class data (black squares in Fig.~\ref{fig: Mixer_CG_Over_Pdc}), using an exponential fit for the frequency range $0.9 \leq f \leq 140~\text{GHz}$, with an R-squared value of 0.5, as shown in Fig.~\ref{fig: Mixer_CG_Over_Pdc} (top). Subsequently, Fig.~\ref{fig: Mixer_CG_Over_Pdc} (bottom) illustrates the \(\mathrm{P_{DC}}\) model for mixers as a function of frequency, with \(\mathrm{P_{IF}}\) set at -5 dBm, demonstrating their low power consumption. However, as frequencies increase, power consumption rises slightly, though this increase remains negligible for practical purposes.

\begin{figure}[t]
\centering
\includegraphics[width=85mm]{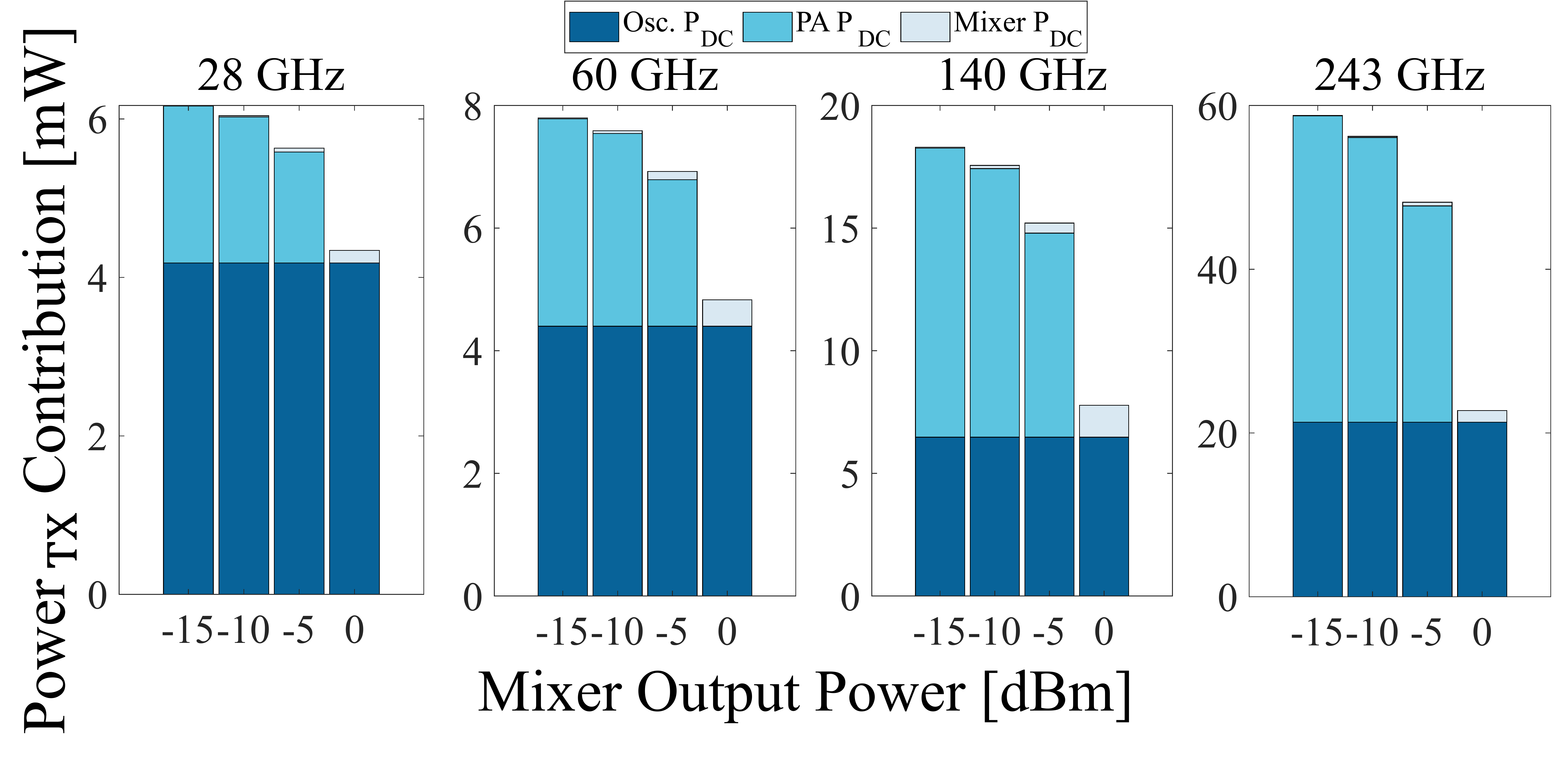} 
\includegraphics[width=85mm]{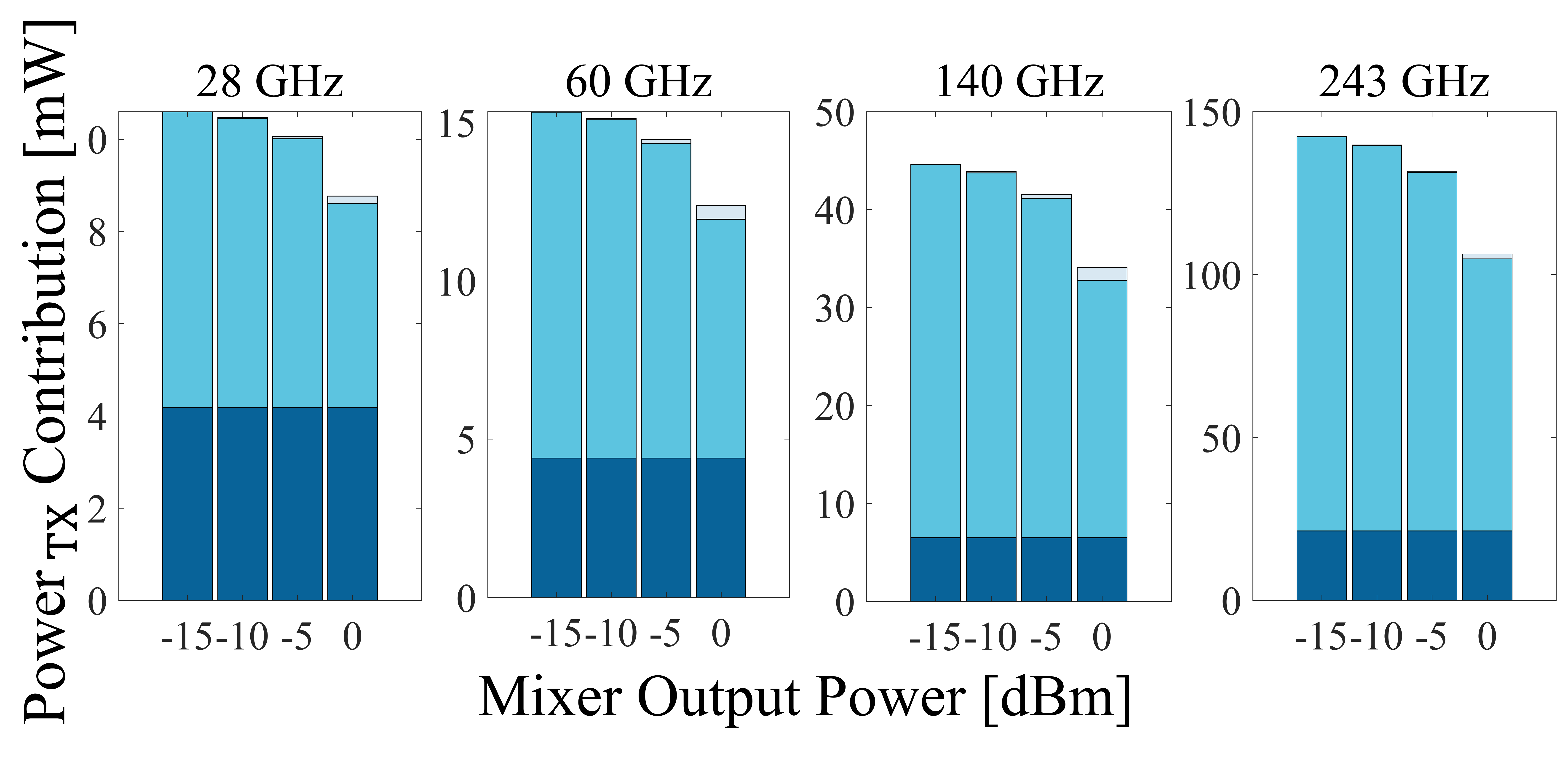} 
\caption{TX front-end DC power contribution for different mixer \(\mathrm{P_{out}}\) levels, with \(\mathrm{P_{IF}} = -5\)~dBm. Top: \(\mathrm{P_{out,PA}}\) and \(\mathrm{P_{out,Osc}}\) = 0~dBm. Bottom: \(\mathrm{P_{out,PA}}\) = 5~dBm, \(\mathrm{P_{out,Osc}}\) = 0~dBm.}
\label{fig: TX_Pdc_Model_5dBm_PA_and_0dBm_PA}
\vspace{-\baselineskip}
\end{figure}

\subsection{RF Front-End Behavioral Power Model for WNoC}

By modeling the individual RF front-end components, developing a comprehensive model for the TX chain becomes feasible. Each sub-block has been analyzed in detail, modeling power consumption as a function of operating frequency for four mixer \(\mathrm{P_{out}}\) levels ([-15, -10, -5, 0] dBm), with a fixed oscillator \(\mathrm{P_{RF, out}}\) of 0~dBm. Also, as an advantage of our model, to make the visualization more practical, we focused on four key frequencies according to the state of the art (Fig.~\ref{fig: TX_Pdc_Model_5dBm_PA_and_0dBm_PA}).

This comprehensive model serves as a valuable tool for calculating TX power consumption, providing remarkable insights that enable designers to optimize WNoC systems. In low-power operation scenarios, with \(\mathrm{P_{out,~PA}}\) set to 0~dBm, the oscillator primarily draws power, accounting for a significant portion of total energy consumption at lower frequencies (see Fig.~\ref{fig: TX_Pdc_Model_5dBm_PA_and_0dBm_PA}, top). In contrast, as frequency increases, even in low-power PA scenarios, the PA becomes the dominant power-consuming component, outpacing other elements. When \(\mathrm{P_{out,~PA}}\) is increased to 5~dBm and frequency rises, the PA's power consumption becomes significantly more pronounced, as seen in Fig.~\ref{fig: TX_Pdc_Model_5dBm_PA_and_0dBm_PA} (bottom). Furthermore, by adjusting the mixer's \(\mathrm{P_{IF}}\) in accordance with our model, various operating configurations can be managed, highlighting a key advantage of this approach. It is noteworthy that the mixer’s contribution at 243~GHz is estimated through extrapolation.

Moreover, since this model was developed based on published works and extensive surveys, and given the absence of global standards for WNoC architecture, its applicability to WNoC systems implicitly validates its feasibility.

The developed \( P_{\text{DC}} \) model for RF front-end CMOS TXs in WNoC systems demonstrates an exponential dependency on frequency. By fine-tuning parameters according to application-specific demands, such as \( P_{\text{in/out,~PA}} \), a precise model of the entire TX chain can be achieved in WNoC systems. This approach is part of an evolving trend in the behavioral modeling of transceivers, which can be further enhanced by incorporating detailed channel models for improved accuracy. Subsequent research will consider factors such as data rate and core area, contributing to the refinement of as yet unspecified parameters in WNoC systems.

\section{Conclusion}\label{Section: Conclusion}

This paper presents a survey-oriented method to model power consumption in WNoC architectures, concentrating on CMOS RF front-end components across different operating frequencies. By adjusting the RF, IF, and LO power levels and calculating the power consumption of sub-blocks, the model provides a flexible method for quantifying power consumption in various scenarios. Results indicate that for low-power PAs, oscillator power usage dominates at frequencies below 60 GHz (exceeding 55\% contribution), while PA power consumption escalates exponentially for low-power scenarios above 60 GHz and consistently dominates in high-power cases during all scenarios and frequencies (exceeding 65\% contribution). Applying this technique to the receiver chain could develop a comprehensive transceiver model, allowing for more precise power calculation and optimization in future designs.

\section*{Acknowledgment}

The authors acknowledge support from the European Commission through grants HORIZON-EIC-2022-PATHFINDEROPEN-01-101099697 (``QUADRATURE''), HORIZON-ERC-2021-101042080 (``WINC''), and HORIZON-MSCA-2022-PF-101062272 (``ENSPEC6G''), as well as from TÜBİTAK through grant 121C073.

\bibliographystyle{IEEEtran}
\bibliography{References}

\end{document}

%% file: modify_IEEEtran.tex
\makeatletter

\def\@maketitle{\newpage
\bgroup\par\addvspace{0.5\baselineskip}\centering%
\ifCLASSOPTIONtechnote
   {\bfseries\large\@IEEEcompsoconly{\sffamily}\@title\par}\vskip 1.3em{\lineskip .5em\@IEEEcompsoconly{\sffamily}\@author
   \@IEEEspecialpapernotice\par{\@IEEEcompsoconly{\vskip 1.5em\relax
   \@IEEEtitleabstractindextextbox{\@IEEEtitleabstractindextext}\par
   \hfill\@IEEEcompsocdiamondline\hfill\hbox{}\par}}}\relax
\else
   \vskip0.2em{\EuMWtitlesize\ifCLASSOPTIONtransmag\bfseries\LARGE\fi\@IEEEcompsoconly{\sffamily}\@IEEEcompsocconfonly{\normalfont\normalsize\vskip 2\@IEEEnormalsizeunitybaselineskip
   \bfseries\Large}\@title\par}\vskip1.0em\par
   \ifCLASSOPTIONconference%
      {\@IEEEspecialpapernotice\mbox{}\vskip\@IEEEauthorblockconfadjspace%
       \mbox{}\hfill\begin{@IEEEauthorhalign}\@author\end{@IEEEauthorhalign}\hfill\mbox{}\par}\relax
   \else
      \ifCLASSOPTIONpeerreviewca
         {\@IEEEcompsoconly{\sffamily}\@IEEEspecialpapernotice\mbox{}\vskip\@IEEEauthorblockconfadjspace%
          \mbox{}\hfill\begin{@IEEEauthorhalign}\@author\end{@IEEEauthorhalign}\hfill\mbox{}\par
          {\@IEEEcompsoconly{\vskip 1.5em\relax
           \@IEEEtitleabstractindextextbox{\@IEEEtitleabstractindextext}\par\hfill
           \@IEEEcompsocdiamondline\hfill\hbox{}\par}}}\relax
      \else
         \ifCLASSOPTIONtransmag
           {\@IEEEspecialpapernotice\mbox{}\vskip\@IEEEauthorblockconfadjspace%
            \mbox{}\hfill\begin{@IEEEauthorhalign}\@author\end{@IEEEauthorhalign}\hfill\mbox{}\par
           {\vspace{0.5\baselineskip}\relax\@IEEEtitleabstractindextextbox{\@IEEEtitleabstractindextext}\vspace{-1\baselineskip}\par}}\relax
         \else
           {\lineskip.5em\@IEEEcompsoconly{\sffamily}\sublargesize\@author\@IEEEspecialpapernotice\par
           {\@IEEEcompsoconly{\vskip 1.5em\relax
            \@IEEEtitleabstractindextextbox{\@IEEEtitleabstractindextext}\par\hfill
            \@IEEEcompsocdiamondline\hfill\hbox{}\par}}}\relax
         \fi
      \fi
   \fi
\fi\par\addvspace{0.0\baselineskip}\egroup}

\def\EuMWtitlesize{\@setfontsize{\EuMWtitlesize}{24}{24pt}}
\def\EuMWauthorsize{\@setfontsize{\EuMWauthorsize}{11}{11pt}}
\def\EuMWaffilsize{\@setfontsize{\EuMWaffilsize}{10}{10pt}}
\def\EuMWcaptionsize{\@setfontsize{\EuMWcaptionsize}{9}{10pt}}
\def\EuMWbibsize{\@setfontsize{\EuMWbibsize}{8}{10pt}}

\def\@IEEEauthorblockNstyle{\EuMWauthorsize\@IEEEcompsocnotconfonly{\sffamily}\@IEEEcompsocconfonly{\large}}
\def\@IEEEauthorblockAstyle{\EuMWaffilsize\@IEEEcompsocnotconfonly{\sffamily}\@IEEEcompsocconfonly{\itshape}\@IEEEcompsocconfonly{\large}}
\def\@IEEEauthordefaulttextstyle{\EuMWauthorsize\@IEEEcompsocnotconfonly{\sffamily}\sublargesize}

\def\thebibliography#1{\section*{\refname}%
    \addcontentsline{toc}{section}{\refname}%
    \EuMWbibsize\@IEEEcompsocconfonly{\small}\vskip 0.3\baselineskip plus 0.1\baselineskip minus 0.1\baselineskip
    \list{\@biblabel{\@arabic\c@enumiv}}%
    {\settowidth\labelwidth{\@biblabel{#1}}%
    \leftmargin\labelwidth
    \advance\leftmargin\labelsep\relax
    \itemsep \IEEEbibitemsep\relax
    \usecounter{enumiv}%
    \let\p@enumiv\@empty
    \renewcommand\theenumiv{\@arabic\c@enumiv}}%
    \let\@IEEElatexbibitem\bibitem%
    \def\bibitem{\@IEEEbibitemprefix\@IEEElatexbibitem}%
\def\newblock{\hskip .11em plus .33em minus .07em}%
\ifCLASSOPTIONtechnote\sloppy\clubpenalty4000\widowpenalty4000\interlinepenalty100%
\else\sloppy\clubpenalty4000\widowpenalty4000\interlinepenalty500\fi%
    \sfcode`\.=1000\relax}

\long\def\@makecaption#1#2{%
\ifx\@captype\@IEEEtablestring%
\par\@IEEEtabletopskipstrut
\else
\@IEEEfigurecaptionsepspace
\fi
\setbox\@tempboxa\hbox{\normalfont\footnotesize {#1.}\nobreakspace\nobreakspace #2}%
\ifdim \wd\@tempboxa >\hsize%
\setbox\@tempboxa\hbox{\normalfont\footnotesize {#1.}\nobreakspace\nobreakspace}%
\parbox[t]{\hsize}{\normalfont\footnotesize\noindent\unhbox\@tempboxa#2}%
\else
\ifCLASSOPTIONconference \hbox to\hsize{\normalfont\footnotesize\hfil\box\@tempboxa\hfil}%
\else \hbox to\hsize{\normalfont\footnotesize\box\@tempboxa\hfil}%
\fi\fi
\ifx\@captype\@IEEEtablestring%
\@IEEEtablecaptionsepspace
\else
\fi}

\newlength\tablecaptiontotableskip
\newlength\figuretocaptionskip
\def\@IEEEfigurecaptionsepspace{\vskip\figuretocaptionskip\relax}%
\def\@IEEEtablecaptionsepspace{\vskip\tablecaptiontotableskip\relax}%

\def\abstract{\normalfont%
\@IEEEabskeysecsize\bfseries\textit{\abstractname}\,\bfseries\textit{---}\,%
\@IEEEgobbleleadPARNLSP}%

\def\IEEEkeywords{\normalfont%
\@IEEEabskeysecsize\bfseries\textit{\IEEEkeywordsname}\,\bfseries\textit{---}\,%
\@IEEEgobbleleadPARNLSP}%
\def\endIEEEkeywords{\relax\vspace{0.67ex}%
\par\if@twocolumn\else\endquotation\fi%
\normalsize\normalfont}%

\DeclareRobustCommand*{\EuMWauthorrefmark}[1]{\raisebox{0pt}[0pt][0pt]{\textsuperscript{#1}}}%
\def\@IEEEauthorblockNtopspace{0ex}
\def\@IEEEauthorblockAtopspace{1mm}
\def\tablename{Table}
\def\thetable{\arabic{table}}
\def\IEEEkeywordsname{Keywords}
\def\subsubsection{\@startsection{subsubsection}{3}{\z@}{1.5ex plus 1.5ex minus 0.5ex}%
{0.7ex plus .5ex minus 0ex}{\normalfont\normalsize\itshape}}%
\newlength{\CPheadmatchindent}%
\def\@seccntformat#1{\hbox to\CPheadmatchindent{\csname the#1dis\endcsname}\hskip 0.1em \relax}
\IEEEilabelindentA \parindent
\IEEEilabelindent \IEEEilabelindentA
\IEEEelabelindent \parindent
\IEEEdlabelindent \parindent
\IEEElabelindent \parindent
\makeatother